\documentclass[aps,amsmath,notitlepage,twocolumn,amssymb,prl,longbibliography]{revtex4-1}

\usepackage{graphicx}
\usepackage{dcolumn}
\usepackage{bm}
\usepackage[colorlinks=true,linkcolor = blue,citecolor=blue,urlcolor=blue,anchorcolor = blue]{hyperref}
\usepackage{wasysym}
\usepackage{stmaryrd}
\usepackage{verbatim}
\usepackage{subfigure}
\usepackage{amsmath}
\usepackage{times}
\usepackage[version=4]{mhchem}
\usepackage{braket}
\usepackage{siunitx}
\usepackage{siunitx}
\usepackage{amssymb}
\usepackage{float}
\usepackage{threeparttable}    
\newcommand{\RNum}[1]{\uppercase\expandafter{\romannumeral #1\relax}}

\begin{document}

\title{Low-energy Spin Dynamics of Quantum Spin Liquid Candidate \ce{NaYbSe2}}

\author{Zheng\,Zhang$^{1,5}$}
\author{Jianshu\,Li$^{1}$}
\author{Mingtai\,Xie$^{2}$}
\author{Weizhen\,Zhuo$^{2}$}
\author{D.T.\,Adroja$^{3,6}$}
\author{Peter J.\,Baker$^{3}$}
\author{T.G.\,Perring$^{3}$}
\author{Anmin\,Zhang$^{2}$}
\author{Feng\,Jin$^{1}$}
\author{Jianting\,Ji$^{1}$}
\author{Xiaoqun\,Wang$^{4}$}
\author{Jie\,Ma$^{4}$}
\email[e-mail:]{jma3@sjtu.edu.cn}
\author{Qingming\,Zhang$^{2,1}$}
\email[e-mail:]{qmzhang@ruc.edu.cn}

\affiliation{$^{1}$Beijing National Laboratory for Condensed Matter Physics, Institute of Physics, Chinese Academy of Sciences, Beijing 100190, China}
\affiliation{$^{2}$School of Physical Science and Technology, Lanzhou University, Lanzhou 730000, China}
\affiliation{$^{3}$ISIS Neutron and Muon Facility, STFC Rutherford Appleton Laboratory, Chilton, Didcot Oxon, OX11 0QX, United Kingdom}
\affiliation{$^{4}$Department of Physics and Astronomy, Shanghai Jiao Tong University, Shanghai 200240, China}
\affiliation{$^{5}$Department of Physics, Renmin University of China, Beijing 100872, China}
\affiliation{$^{6}$Highly Correlated Matter Research Group, Physics Department, University of Johannesburg, PO Box 524, Auckland Park 2006, South Africa}

\date{\today}

\begin{abstract}

The family of rare-earth chalcogenides \ce{ARECh2} (A = alkali or monovalent ions, RE = rare earth, and Ch =
O, S, Se, and Te) appears as an inspiring playground for studying quantum spin liquids (QSL). The crucial low-energy spin dynamics remain to be uncovered. By employing muon spin relaxation ($\mu$SR) and zero-field (ZF) AC susceptibility down to 50 mK, we are able to identify the gapless QSL in \ce{NaYbSe2}, a representative member with an effective spin-1/2, and explore its unusual spin dynamics. The ZF $\mu$SR experiments
unambiguously rule out spin ordering or freezing in \ce{NaYbSe2} down to 50 mK, two orders of magnitude smaller than
the exchange coupling energies. The spin relaxation rate, $\lambda$, approaches a constant below 0.3 K, indicating finite
spin excitations be featured by a gapless QSL ground state. This is consistently supported by our AC susceptibility
measurements. The careful analysis of the longitudinal field (LF) $\mu$SR spectra reveals a strong spatial correlation
and a temporal correlation in the spin-disordered ground state, highlighting the unique feature of spin
entanglement in the QSL state. The observations allow us to establish an experimental H-T phase diagram. The
study offers insight into the rich and exotic magnetism of the rare earth family.
\end{abstract}

\maketitle

\emph{Introduction}---Recently revealed triangular lattice rare-earth chalcogenides form a large family of quantum spin liquid (QSL) candidates\cite{liu2018rare,baenitz2018naybs,bordelon2019field,ranjith2019anisotropic}. The strong spin-orbit coupling (SOC) and crystalline electronic field (CEF) excitations generate a rich diversity of magnetic and electronic properties of the family and have been extensively investigated\cite{PhysRevB.101.224427,ding2019gapless,baenitz2018naybs,Zhang2020,dai2020spinon,PhysRevB.102.024424,PhysRevB.101.144432}.

As a representative member of the family, \ce{NaYbSe2} receives increasing research interest, particularly on the experimental side, because of the availability of larger single crystals. In our previous work, we have carried out {inelastic neutron scattering (INS)} and Raman studies of the CEF excitations in \ce{NaYbSe2}, and revealed that \ce{NaYbSe2} has the smallest CEF compared to \ce{NaYbO2} and \ce{NaYbS2}\cite{Zhang2020}. By applying the CEF theory and the mean-field (MF) picture to thermodynamic measurements, we are able to further determine a characteristic temperature of $\sim$ 25 K, below which CEF excitations fade away and intrinsic spin interactions come to dominate low-energy magnetic excitations. This guarantees that at a sufficiently low temperature, one can safely and strictly describe the spin system using an anisotropic spin-1/2 Hamiltonian on a perfect triangular lattice.

The highly anisotropic spin interactions featured by rare-earth magnetic ions, may give rise to nontrivial magnetic properties and ground states. INS experiments have reported the observations of spinon excitations under zero field\cite{bordelon2019field,dai2020spinon}. Several unusual spin-ordered phases can be induced by magnetic fields\cite{bordelon2019field,ranjith2019anisotropic,xing2019field}. More interestingly, unconventional superconductivity in \ce{NaYbSe2} comes into consideration due to its appropriate energy band gap ($\sim$ 1.9 eV) and actually some attempts with high pressure have been carried out\cite{Jia_2020,zhang2020pressure}. All these are ultimately related to spin dynamics in the ground state, which remains to be clearly uncovered because of the limited energy resolution of INS experiments. {The muon spin rotation and relaxation ($\mu$SR)} technique is generally recognized as a powerful and unique tool to probe low-energy spin dynamics, since muons injected into the sample are extremely sensitive to any tiny local internal field {at the muon stopping sites}. {Since selenium ions are the only negatively charged anions in \ce{NaYbSe2}, the implemented $\mu^{+}$ mainly stops near \ce{Se$^{2-}$} and probes local magnetic environment and spin dynamics (see Fig. 1).} 

In this letter, we explore the magnetic ground state of \ce{NaYbSe2} by combining $\mu$SR and zero-field AC susceptibility down to 50 mK.
The ZF and longitudinal-field (LF) $\mu$SR measurements, together with ZF AC susceptibility measurements, unambiguously rule out any spin ordering or freezing in the ground state in \ce{NaYbSe2}. 

The muon spin relaxation rate $\lambda$ is a highly sensitive measure of spin fluctuations and approaches a {constant value of $\sim$ 0.9 $\mu$s$^{-1}$} below 0.3 K, which is far beyond the contribution from the wavefunction of the CEF ground state. The signal of finite spin excitations is also supported by AC susceptibility measurements, and explicitly points to a gapless QSL ground state. The careful analysis of LF $\mu$SR spectra offers more insight into spin dynamics, which suggests a strong spin correlation in both space and time, a signal of spin entanglement unique to QSL. The observations allow to establish an experimental H-T phase diagram. 

\begin{figure}[t]
	\includegraphics[scale=1]{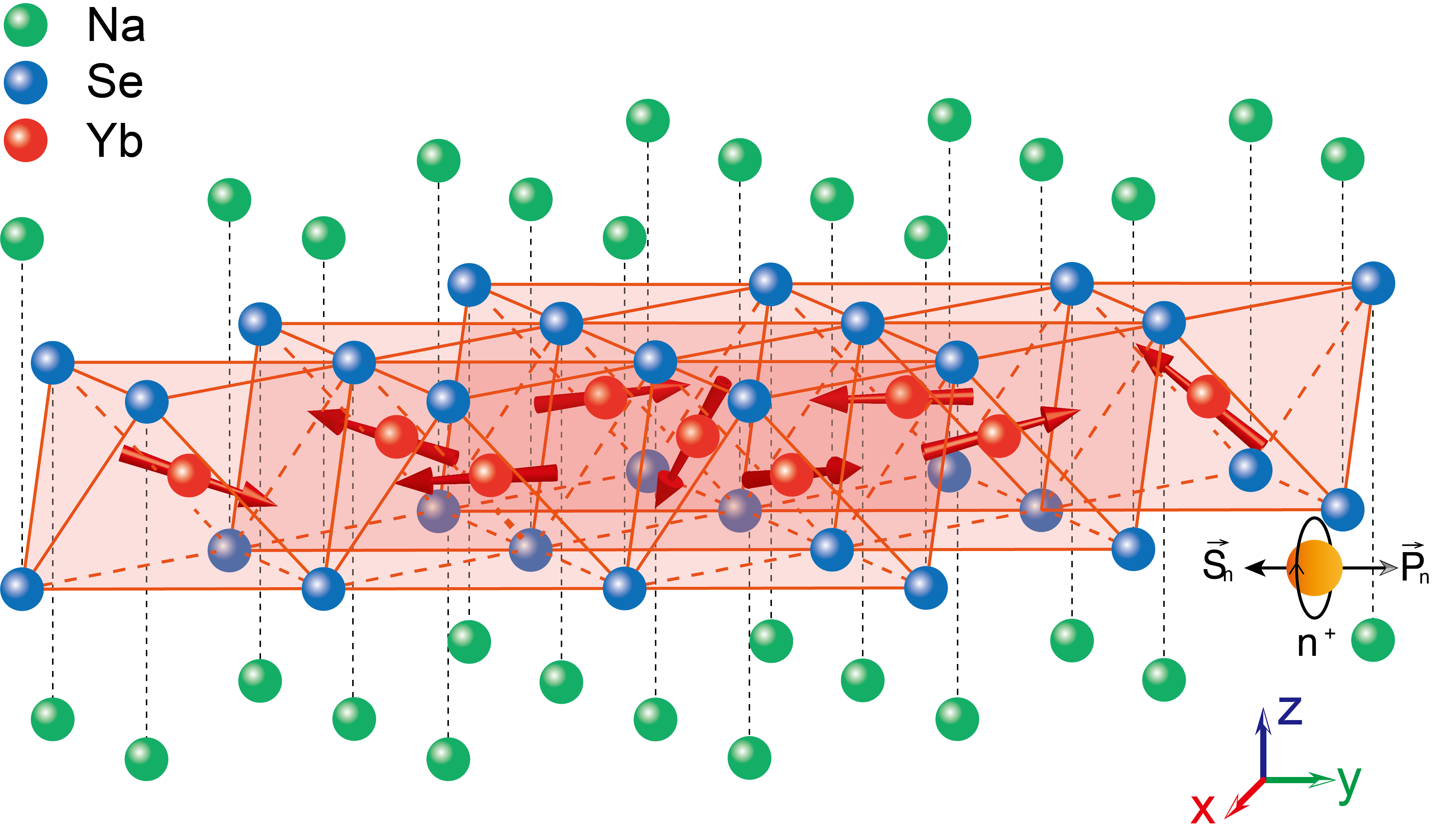}
	\caption{\label{fig:epsart}Crystal structure of \ce{NaYbSe2} and schematic diagram of muon stopping site in \ce{NaYbSe2}. The implanted $\mu^{+}$ mainly stops near \ce{Se^{2-}} and probes local magnetic environment and spin dynamics.}
\end{figure}

\emph{Experiments and numerical methods}---Single crystals of \ce{NaYbSe2} were grown by the \ce{NaCl}-flux method and the polycrystalline samples were synthesized by the \ce{Se}-flux method\cite{liu2018rare, Zhang2020}. The high quality of both single-crystal and polycrystalline samples were confirmed by x-ray diffraction (XRD), Laue x-ray diffraction, and Energy Dispersive X-Ray Spectroscopy (EDX)\cite{Zhang2020}. Single crystals were orientated using Laue x-ray diffraction and fixed on an Ag sample holder.

The $\mu$SR data was collected using the MuSR spectrometer at ISIS facility, Rutherford Appleton Laboratory, United Kingdom. The data in the high-temperature range of 4 $\sim$ 10 K was collected with a He-4 cryostat, while the data in the low-temperature range of 0.05 $\sim$ 4 K was collected with a dilution refrigerator. The $\mu$SR data was analyzed with the open source software package Mantid\cite{Arnold2014}. The data of ZF-$\mu$SR, LF-$\mu$SR at 1 K, and LF-$\mu$SR at 0.1 K are provided by the polycrystalline of \ce{NaYbSe2}; the data of LF-$\mu$SR at 10 K is provided by single crystal of \ce{NaYbSe2}.
For the $\mu$SR data, the effective measurement time starts at t = 0.14 $\mu$s.

The ZF AC susceptibility measurements from 50 mK to 4 K were performed on a dilution refrigeration (DR) system. 

\begin{figure}[t]
	\includegraphics[scale=1]{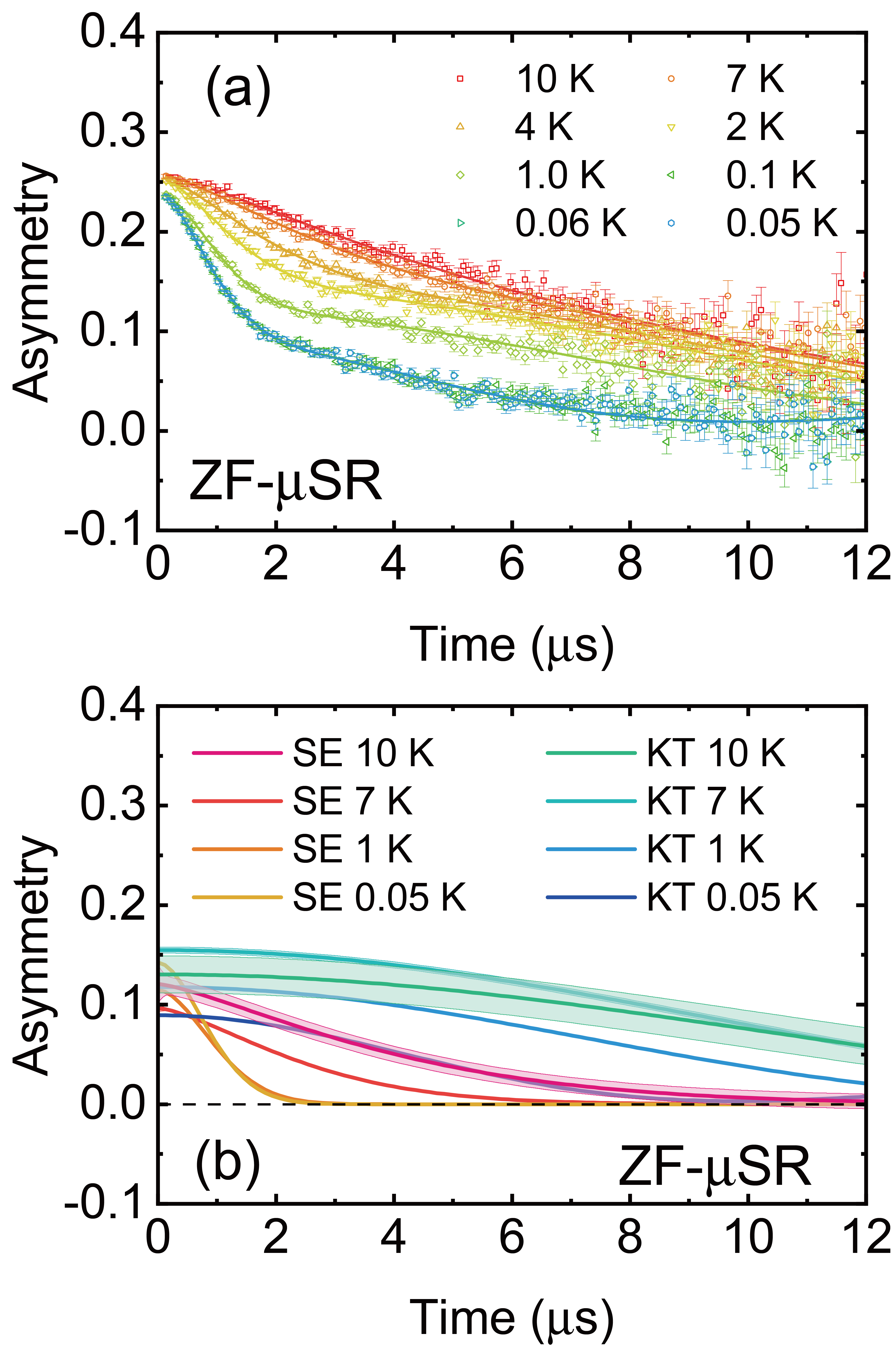}
	\caption{\label{fig:epsart} {ZF $\mu$SR spectra of \ce{NaYbSe2}. (a) ZF-$\mu$SR asymmetry spectra with the fitting curves using formula (1).  (b) The contribution of the stretching exponential (SE) term and Kubo-Toyabe (KT) term obtained from (a) at the selected temperatures.
	The shades accompanying the fitting curves in the (b) are actually the fitting error bars.}
	}
\end{figure}

\begin{figure*}[t]
	\includegraphics[scale=0.9]{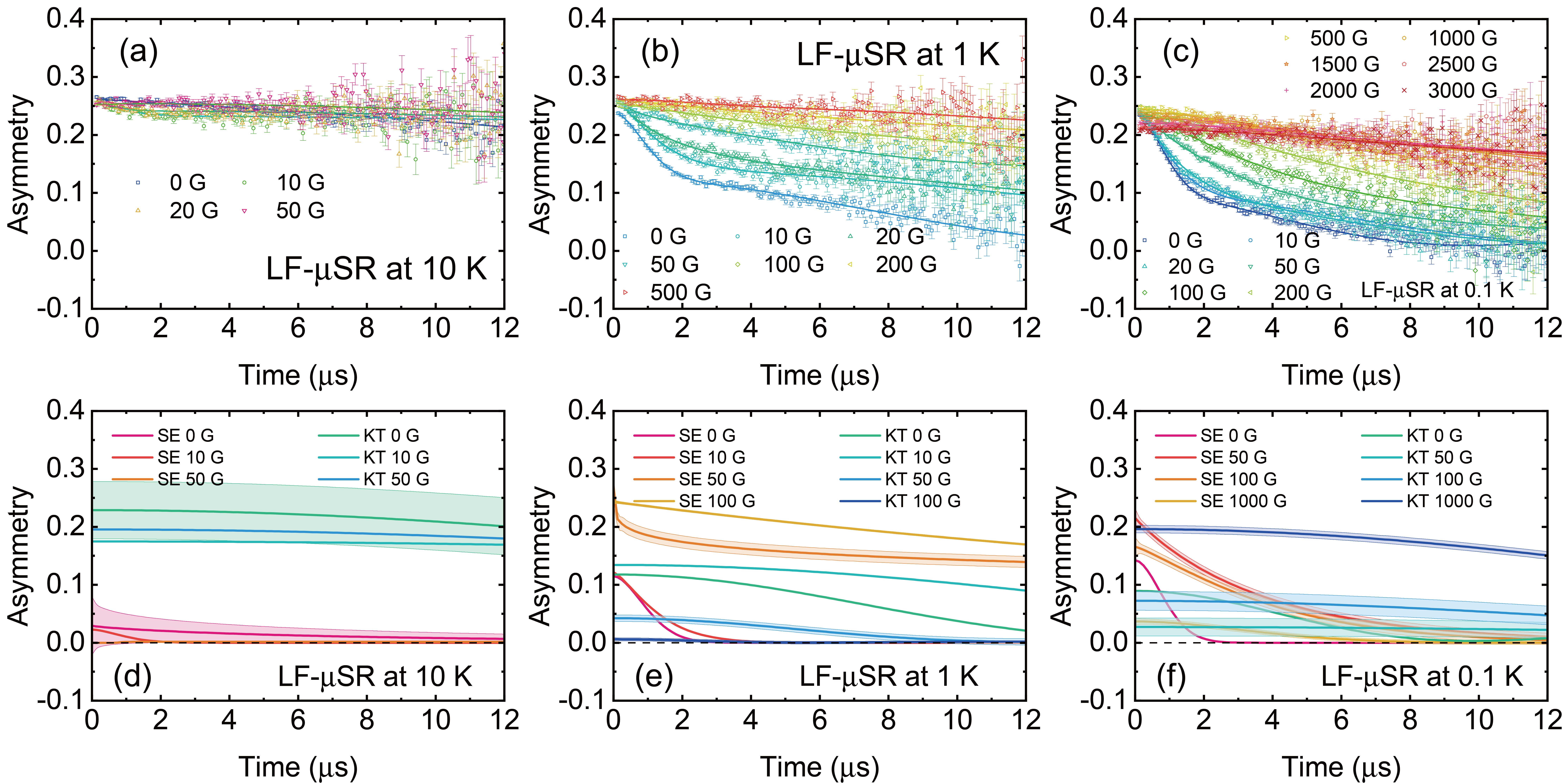}
	\caption{\label{fig:epsart} {LF-$\mu$SR spectra of \ce{NaYbSe2}.  (a), (b), and (c) LF-$\mu$SR asymmetry spectra at 10 K, 1 K, and 0.1 K, with the fittings by formula (1). (f), (g), and (h) The contribution of the stretching exponential (SE)
			term and the Kubo-Toyabe (KT) term corresponding to (d), (e), and (f) at the selected magnetic fields. The shades accompanying the fitting curves in the (d) $\sim$ (f) are actually the fitting error bars.}
	}
\end{figure*}


\emph{Zero-Field (ZF) $\mu$SR}---
The ZF-$\mu$SR asymmetry spectra are analysed with a relatively complete formula written as:
\begin{equation}
	\begin{split}
		& A(t) = A_{0} + A_{SE} + A_{KT} \\
		& A_{SE} = A_{1}e^{-(\lambda t)^{\beta}} \\
		& A_{KT} = A_{2}\left[\frac{2}{3}(1-\sigma^{2}t^{2})e^{-\frac{1}{2}\sigma^{2}t^{2}}+\frac{1}{3}\right]
	\end{split}
\end{equation}
where the first term on the right side, $A_{0}$, is the flat background term originating from the sample holder {and maintains a constant value of 0.0062 during the fitting process}; the second term comes from the stretched exponential relaxation and $A_{1}$ is the initial asymmetry for this component, $\lambda$ the $\mu^{+}$ spin relaxation rate, $\beta$ the stretching exponent, and the third one is called as Kubo-Toyabe term representing  the width of the static field distribution {term and the decay rate $\sigma$ is related to the width of dipolar field distribution} \cite{PhysRevB.20.850,Zhu2020}.

First of all, no oscillation is seen in the ZF-$\mu$SR spectra within the time window up to 12 $\mu$s, even down to 50 mK in Fig. 2(a). Actually, there is no evidence for oscillations over the full measured time window of 32 $\mu$s. This unambiguously rules out the possibility of the magnetically ordered ground state.

Further analysis allows to exclude a global spin freezing in the ground state. (1) The polarization in the ZF-$\mu$SR spectra does not recover to 1/3 of the initial one. This may suggest a small amount of spin-glass-like freezing related to magnetic defects. The similar observations are also reported in \ce{YbMgGaO4}\cite{li2016muon}, \ce{NaYbO2}\cite{ding2019gapless}, and \ce{NaYbS2}\cite{sarkar2019quantum}. 
(2) 
{The stretched exponent $\beta$ fluctuates around 1.8 in the temperature range of 0.3 $\sim$  1 K (see Fig. 4(b))}, which corresponds to a broad peak in heat capacity\cite{Zhang2020a}. This indicates that the spin system establishes a strong spin correlation in the temperature range. The fact that $\beta$ below 0.3 K is close to 1.8 rather than around 1/3, suggests that the spin system develops into a stable quantum spin liquid despite a small amount of spin freezing or magnetic defects\cite{PhysRevB.98.174404}. 
A global spin freezing usually gives a $\beta$ of $\sim$1/3 at the freezing temperature\cite{li2016muon}. {It is an empirical rather than strictly derived value. Qualitatively the small $\beta$ in a spin glass is considered to be related to the distribution of relaxation rates centered at a small value. Experimentally the $\beta$ values of spin glass materials AgMn alloy\cite{PhysRevLett.72.1291}, \ce{Y(Mn_{1-x}Al_{x})_{2}}\cite{PhysRevLett.72.1291, cywinski_spin_1994}, and \ce{La2Cu_{0.25}Co_{0.75}O_{4+\delta}}\cite{PhysRevLett.72.1291} are all close to 1/3 below the spin glass transition temperatures. And, the Monte Carlo simulations result also gives a $\beta$ value of $\sim$ 1/3 below the spin-glass transition temperature\cite{PhysRevB.32.7384}. }
(3) The Kubo-Toyabe term can tell us more. The term smoothly decays with time and does not recover to 1/3 of the initial one within the measurement time (see Fig. 2(b)). It is insufficient to support a spin freezing. The fitting parameter $\sigma$ in the Kubo-Toyabe term increases with lowering temperatures (see Fig. 4(c)). This is related to nuclear magnetic moments rather than spin freezing. The Schottky effect of nuclear magnetic moments becomes remarkable at very low temperatures, and has been observed in the zero-field heat capacity\cite{Zhang2020a}. Expectedly, it is also observed in \ce{NaYbS2}\cite{sarkar2019quantum} and \ce{NaYbSe2}. It should be pointed out that there is no such static field observed in \ce{YbMgGaO4} even below 0.1 K, as the Mg/Ga disorder may smear out the contribution from the nuclear magnetic moment.

Let's turn to the three key parameters describing low-energy spin dynamics, $\lambda$, $\beta$ and $\sigma$. They all go up to a plateau below 0.3 K (Fig. 4), indicating the emergence of a stable magnetically disordered QSL ground state. Among them, the ZF $\mu^{+}$ spin relaxation rate $\lambda$, which probes local spin fluctuations, is a key indicator of the ground-state magnetism. In order to dig into it deeply, we have checked the contribution from the CEF ground state. The spin-orbit coupled CEF ground state in \ce{NaYbSe2} is an effective spin-1/2 Kramers doublet protected by time-reversal symmetry. The careful determination of the CEF parameters by INS and thermodynamic measurements gives rise to the exact wavefunction of the CEF ground state\cite{Zhang2020}. We are able to calculate magnetic susceptibility from 0.01 to 10 K using the CEF ground-state wavefunction\cite{SI}. Not surprisingly, the magnetic susceptibility simply follows the Curie-Weiss law. It is possible that the paramagnetic-like susceptibility develops a plateau at very low temperatures, but only when a sufficiently high magnetic field is applied. This is not the case for the spin relaxation rate $\lambda$ which feels the uniform internal field without any external field. Thus the plateau of $\lambda$ intrinsically comes from finite spin excitations, which evidences a gapless QSL ground state. 
For cross check the picture, we have further measured ZF AC magnetic susceptibility down to 50 mK (Fig. 4(d)). Clearly no magnetic transition is observed here and the absence of frequency dependence also rules out the possibility of spin freezing. More importantly, the susceptibility also approaches a constant at the lowest temperatures, which well supports ZF $\mu$SR results discussed above.
{It is important to note that the plateau of AC susceptibility appears at a lower temperature than that of $\mu$SR. 
One reason is that $\mu$SR measurements probe local magnetism and spin dynamics while AC susceptibility measurements prob the magnetic properties of bulk material. In this sense, the AC susceptibility and spin relaxation rate are not completely equivalent. The spin system gradually enters a QSL state as temperature decreases. When the sign of QSL appears in a local region, it can be caught up by $\mu$SR. AC susceptibility cannot see a stable platform until the system enters a global QSL state. The other reason is that the frequency of muon is up to an order of 10$^{8}$ Hz, while the working frequency of AC susceptibility does not exceed an order of 10$^{4}$ Hz. It means that muon can observe the platform of spin relaxation rate at a relatively higher temperature, while the platform in the channel of AC susceptibility appears normally at a lower temperature.}

\begin{figure}[b]
	\includegraphics[scale=0.34]{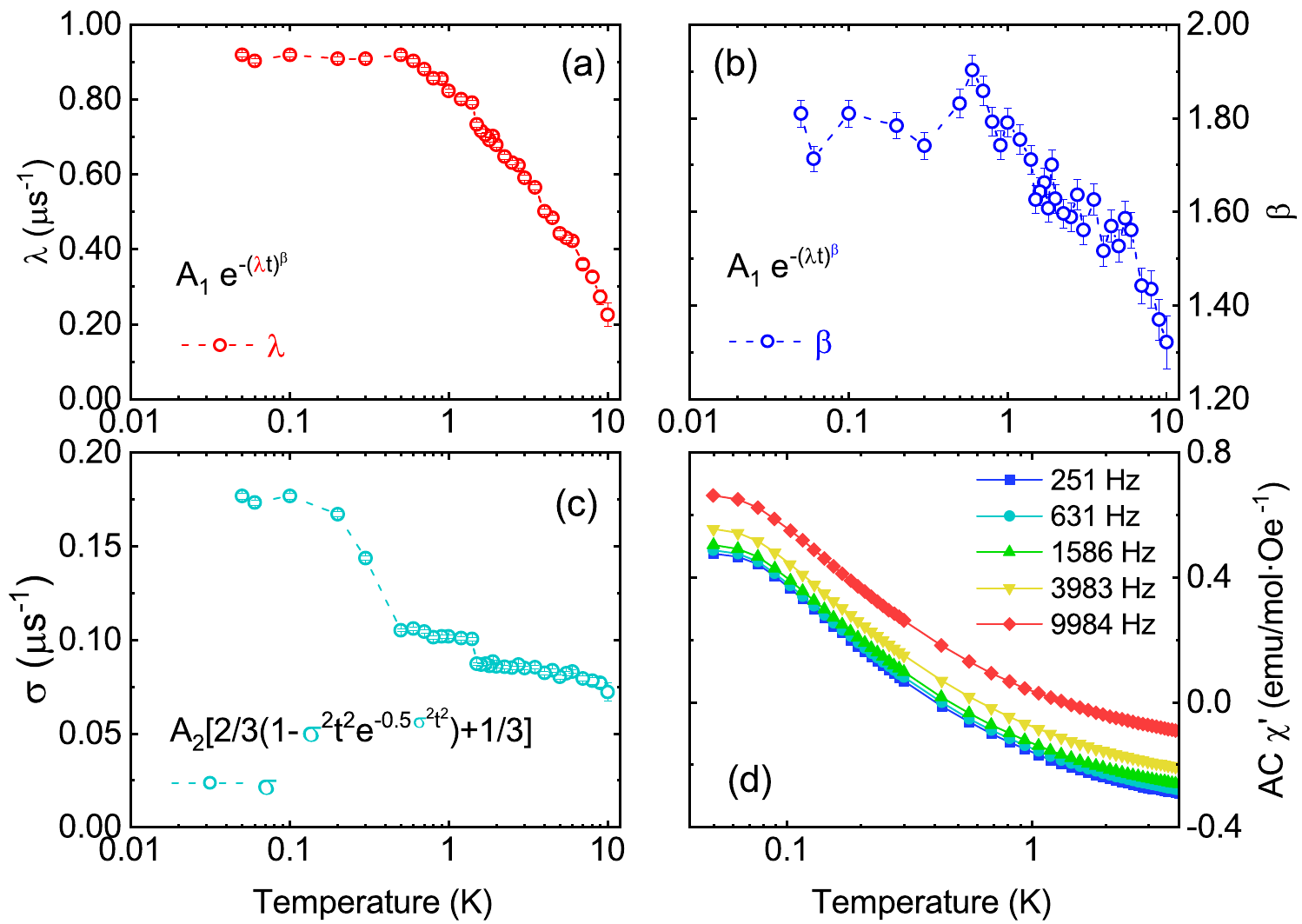}
	\caption{\label{fig:epsart} Temperature evolution of spin dynamics parameters. (a) ZF-$\mu$SR spin relaxation rate $\lambda$. (b) ZF-$\mu$SR stretching exponent $\beta$. (c) Decay rate $\sigma$ in the Kubo-Toyabe term. The three parameters describing spin dynamics are defined in formula (1). (d) ZF AC susceptibility at various frequencies. Magnetic field dependence of spin dynamics. 
}
\end{figure}

\begin{figure}[b]
	\includegraphics[scale=0.21]{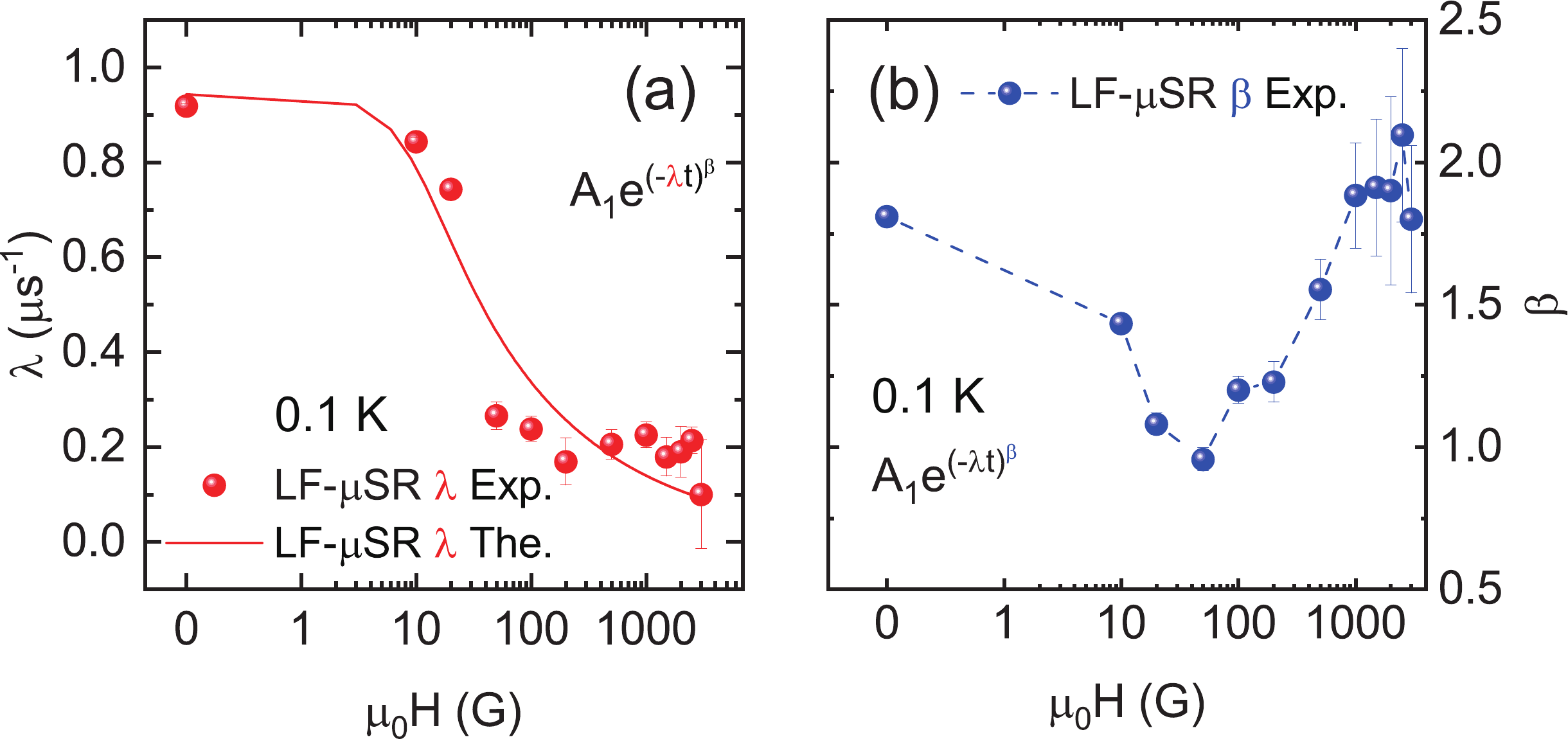}
	\caption{\label{fig:epsart} (a) LF-$\mu$SR spin relaxation rate $\lambda$. The red solid circles are obtained from LF-$\mu$SR spectra, and the red solid line is produced by the expression (3). (b) LF-$\mu$SR stretching exponent $\beta$.}
\end{figure}

\emph{Longitudinal-Field (LF) $\mu$SR}---The LF-$\mu$SR experiments allow to analyze the static and non-static fields in \ce{NaYbSe2}. The stretched exponential term and the Kubo-Toyabe term in the above fitting formula are related to non-static uniform local fields (or non-static fields)\cite{ohira_spin_2006}  and static fields, and reflect spin fluctuations in the QSL phase and the contribution from the nuclear magnetic moment\cite{PhysRevB.71.092501}, respectively. 

At 10 K, the application of a tiny magnetic field of $\sim$ 10 G can polarize the Kubo-Toyabe term to almost a straight line (Fig. 3(a) and Fig. 3(d)), which means there is little static field at this temperature. By lowering the temperatures to 1 K, one can see that the zero-field $\mu$SR asymmetry spectrum shows a weak but clear upturning (Fig. 3(b)), which comes from the Kubo-Toyabe term. In other words, a small but finite
static field appears around 1 K. The static field is suppressed by a magnetic field of $\sim$ 50 G. This suggests that the weak static field in \ce{NaYbSe2} should be related to nuclear magnetic moment. The Kubo-Toyabe term at 0.1 K is similar to what we have seen at 1 K.  {In \ce{NaYbSe2}, $^{23}$Na and $^{79}$Se have non-zero nuclear magnetic moments. The nuclear spin and magnetic moment for $^{23}$Na are 3/2 and 2.21 $\mu_{N}$, and 7/2 and -1.02 $\mu_{N}$ for $^{79}$Se (refers to the nuclear magneton), respectively. Generally a positively charged $\mu^{+}$ stops near \ce{Se^{2-}} rather than \ce{Na^{+}}, and hence probes the local magnetic environment and spin dynamics around $^{79}$Se.}

Compared to the static field, the non-static internal field described by the stretched exponential term is more interesting and informative. {The stretched exponential term is closely related to quantum fluctuations. We use the steady magnetic fields to carry out LF-$\mu$SR measurements.} A clear contribution of the stretched exponential term (or non-static internal field) appears at 10 K, but a weak magnetic field ($\sim$ 10 G) can significantly suppress it (see orange red line in Fig. 3(d)) over the measurement time. At 1 K and 0.1 K, the magnetic fields needed increase to $\sim$50 G and $\sim$1000 G, respectively (Fig. 3(b) (e) (c) (f)). This means that the non-static field (or intrinsic spin fluctuations) related to anisotropic spin-exchange interactions becomes stronger at lower temperature. It reflects that there exists a strong and intrinsic spatial spin correlation and fluctuation at low temperatures in \ce{NaYbSe2}. {The strong spatial and temporal (see below) spin correlations are driven by strong quantum rather than thermal fluctuations in the spin-entangled QSL state.} The similar observation has been reported in some other QSL candidates, such as \ce{Ce2Zr2O7}\cite{gao_experimental_2019} and \ce{NaYbS2}\cite{sarkar2019quantum}. {At different temperatures, the contribution of the Kubo-Toyabe term and the stretched exponential term to the LF-$\mu$SR spectra with the change of magnetic fields can be found in the supplementary materials\cite{SI}.}


\emph{Spin Dynamics}--- Let's turn to the analysis on spin dynamics. The spin relaxation rate $\lambda$ at higher temperatures can be written as\cite{li2016muon},
\begin{equation}
	\lambda (T > 10 K, H) = \frac{2\Delta^{2}\nu}{\nu^{2} + (\mu_{0}H\gamma_{\mu})^{2}}
\end{equation}
where $\lambda$ approaches a temperature-independent value of $\sim$ {0.2} $\mu s^{-1}$ above 10 K (Fig. 4(a)); $\Delta$ is the fluctuating component of the field at the $\mu^{+}$ site perpendicular to its initial polarization, which describes the distribution width of local magnetic fields; $\nu$ is the spin fluctuation frequency; $\gamma_{\mu}$ = 135.5 MHz/T is the $\mu^{+}$ gyromagnetic ratio; and $\mu_0$ is the magnetic permeability in vacuum. The fluctuation frequency at $\sim$ 10 K can be estimate using $\nu=\sqrt{z}J/h$, where the coordination number $z$ is 6 which refers to the number of the nearest \ce{Se^{2-}} neighbors coordinated to a central \ce{Yb^{3+}} magnetic ion. The spin exchange energies are given by analyzing thermodynamic measurements\cite{Zhang2020a}. The calculated results are listed in Table I. The estimated distribution width of local magnetic fields, $\Delta \ll \nu$, confirms the limit of fast spin fluctuation.

\begin{table*}[!ht]
	\caption{Fluctuations of the local fields at 10 K}\label{tab:tablenotes}
	\centering
	\begin{threeparttable}          
		\begin{ruledtabular}
			\begin{tabular}{cccccc}
			Spin interactions & & Fluctuation frequencies & & Distribution widths &\\
			\hline
			$|J_{zz}|$ $^{*}$  & 0.95 K & $\nu_{zz}$ & 2.42 $\times$ 10$^{10}$ Hz & $\Delta_{zz}$ & 5.23 $\times$ 10$^{7}$ Hz\\
			\hline
			$|J_{+-}|$ $^{**}$ & 3.27 K & $\nu_{+-}$ & 8.34 $\times$ 10$^{10}$ Hz& $\Delta_{+-}$ & 9.70 $\times$ 10$^{7}$ Hz\\
			\hline
			$|J_{avg}|$ $^{***}$ & 2.50 K & $\nu_{avg}$ & 6.37 $\times$ 10$^{10}$ Hz& $\Delta_{avg}$ & 8.47 $\times$ 10$^{7}$ Hz\\
			\end{tabular}
		\end{ruledtabular}
	
		\begin{tablenotes}    
			\footnotesize              
			\item[*] The spin-exchange interaction along the c-axis.          
			\item[**] The spin-exchange interaction in the ab-plane.        
			\item[***] $|J_{avg}| = \frac{1}{3}|J_{zz}| + \frac{2}{3}|J_{\pm}|$
		\end{tablenotes}           
	\end{threeparttable}       
\end{table*}

Based on the above results and the following expression of the $\mu^{+}$ spin relaxation rate\cite{PhysRevB.64.054403,PhysRevB.18.3026}, we can further analyze the dynamic spin correlation at low temperatures. The expression has been adopted to analyze spin dynamic characteristics of quantum spin liquid candidates, such as \ce{YbMgGaO4}\cite{li2016muon} and \ce{NaYbS2}\cite{sarkar2019quantum}.

\begin{equation}
	\lambda (H) = 2\Delta^{2}\tau^{x} \int_{0}^{\infty} t^{-x}exp(-\nu t)cos(2\pi \mu_{0} \gamma_{\mu} H t)dt
\end{equation}
where t is time; $\tau$ is the early time cutoff; $\Delta$,  $\gamma_{\nu}$ and $\nu$ have been defined in (2); And $x$ is defined as a critical exponent\cite{RevModPhys.49.435,li2016muon}. The simulation using this expression, presented in Fig. 5(a), gives $x = 0.62(7)$ and $\nu = 8.27(6) \times 10 ^{5}$ Hz. It should be noted that the temporal spin correlation function, $S(t) \sim (\tau / t) ^ {x} exp(-\nu t)$, shows an exponential form multiplied by a power law, rather than the pure exponential one of $S(t) \sim exp(-\nu t)$ at high temperatures\cite{li2016muon}. $\nu$ is much smaller at low temperatures ($\sim$ 0.1 K) compared to $\nu \sim 10 ^ {10}$ Hz at high temperatures ($\sim$ 10 K). 
{The longer spin fluctuation can only be driven by the quantum entanglement of spins at low temperatures, which is similar to the LF-$\mu$SR measurement results of QSL candidates \ce{NaYbS2}\cite{sarkar2019quantum} and \ce{YbMgGaO4}\cite{li2016muon}.}
{The stretching exponent, $\beta$, reflects the stability of the ground state under magnetic fields.
Below 200 G, the $\beta$ values from the ZF-$\mu$SR spectra show a V-shape with magnetic fields, but it is always greater than 1. This is a clear sign ruling out spin freezing which has a close to 1/3 as mentioned above. This suggests a fast spin dynamics which remains under a magnetic field less than 200 G. 
And no signal related to a magnetically ordered state was observed in the range of 0 $\sim$ 3000 G. 
In this sense, the LF-$\mu$SR data at 0.1 K allow to infer that the QSL ground state is robust against external magnetic field.
}

It should be highlighted here that the above features extracted from the analysis on LF-$\mu$SR spectra, are unique to a QSL state and contrarily are absent in spin glass or magnetically ordered materials. For instance, in spin-glass compounds \ce{CuMn} alloy\cite{PhysRevB.31.546} and \ce{AgMn} alloy\cite{PhysRevLett.77.1386}, the spin relaxation rate $\lambda$ drops rapidly below the spin freezing temperature, which means that there is almost no temporal spin correlation when entering the spin freezing phase. For the typical magnetically ordered materials, such as transition metal magnetic materials \ce{RFeAsO} (R = La, Ce, Pr, and Sm)\cite{PhysRevB.80.094524} and rare-earth magnetic material \ce{CeCoGe3}\cite{PhysRevB.88.134416} and triangular spin lattice \ce{Na}$_{x}$\ce{CoO2}$\cdot y$\ce{H2O}\cite{PhysRevLett.92.257007}, the temporal spin correlation gradually approaches 0 below the phase transition temperature. 

\begin{figure}
	\includegraphics[scale=0.9]{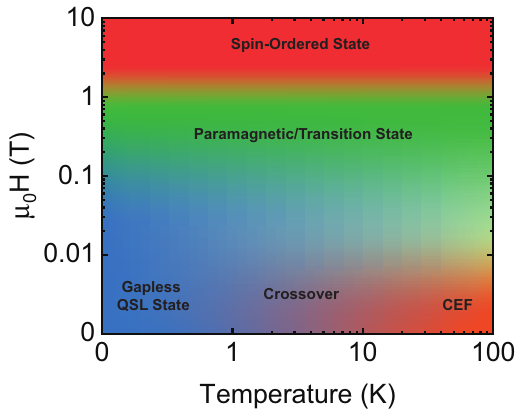}
	\caption{\label{fig:epsart} {Experimental H-T phase diagram of \ce{NaYbSe2} based on the data in this paper and neutron and thermodynamics data from Refs. \cite{Zhang2020} and \cite{Zhang2020a}} . "CEF" refers to the phase in which CEF excitations play a dominant role. "Spin-Ordered State" refers to the magnetically ordered state induced by magnetic fields. 
	}
\end{figure}

\emph{Phase diagram}---The $\mu$SR results identify a gapless QSL ground state in \ce{NaYbSe2} and allow to establish an experimental H-T phase diagram (Fig. 6), 
{in combination of our study of the CEF excitations of \ce{NaYbSe2}\cite{Zhu2020} and the measurements of the magnetic susceptibility, magnetization, and heat capacity of \ce{NaYbSe2} at finite temperatures\cite{Zhang2020a}.}
In the region of T $\textless$ 0.3 K and H $\textless$ 200 G, the spin system is in the gapless quantum spin liquid phase. A weak static field contributed by nuclear moment little affects the ground state, and a weak applied magnetic field ($\sim$ 50 G) can offset the static field. In particular, the spin relaxation rate $\lambda$ given by the ZF-$\mu$SR measurements, basically remains constant ($\sim$ 0.9 $\mu s^{-1}$) in this temperature range, exactly similar to the experimental observation of a gapless quantum spin liquid in \ce{YbMgGaO4}\cite{li2016muon}. 

Considering the phase diagram along the axis of magnetic field. A magnetic field of $\sim$ 1000 G can suppress the non-static uniform field and hence destroy the QSL state. But there is no sign of phase transition around this magnetic field, as we have seen from the thermodynamic results under the same conditions\cite{liu2018rare}. It may mean that the spin system enters a paramagnetic state or transition state in the magnetic field region. A larger magnetic field ($\textgreater$ 3 T) induces magnetically ordered phases, which have been observed and studied in \ce{NaYbO2}\cite{bordelon2019field} and \ce{NaYbS2}\cite{ma2020spin}.

As a function of temperature, one can see another crossover region of 1 $\sim$ 10 K. In this temperature range, spin interactions are almost the same order of magnitude as temperature and the parameters given by the ZF-$\mu$SR spectra have been changed. 
{Meanwhile, we can observe a broad heat capacity peak contributed by spin-exchange interactions near 1 K in the ZF heat capacity of \ce{NaYbSe2}\cite{Zhang2020a}.}
{In the region of higher temperatures ($\textgreater$ 25 K), the susceptibility and heat capacity calculated by using the CEF parameters determined by INS\cite{Zhang2020} are consistent with the experimental results\cite{Zhang2020a}. It suggests that the magnetism in the temperature range is dominated by the CEF excitations.}

\emph{Conclusion}---We have explored the spin ground state and its spin dynamics in \ce{NaYbSe2} by combining $\mu$SR and ZF AC susceptibility. The zero-field $\mu$SR experiments demonstrate that there is no spin ordering or freezing even down to 0.05 K. The spin relaxation rate $\lambda$ stays at a constant of $\sim$ 0.9 $\mu$s$^{-1}$ below 0.3 K, which demonstrates the existence of finite spin excitations in the ground state and hence evidences a gapless QSL ground state. The ZF AC susceptibility also approaches a plateau at very low temperatures and further supports the picture of the gapless QSL ground state. The careful analysis of LF-$\mu$SR spectra offers many characteristics of spin dynamics in \ce{NaYbSe2}, particularly revealing a strong spin correlation in the ground state not only in space but also in time which 
{can be considered as an indication of spin entanglement in QSL.}
The QSL ground state has been also observed in the two brothers of \ce{NaYbSe2}, \ce{NaYbO2}\cite{bordelon2019field,ding2019gapless} and \ce{NaYbS2}\cite{sarkar2019quantum}. This experimentally generalizes the material system as an inspiring platform for studying QSL. 

The study sheds light on the unconventional magnetic properties of the rare-earth family and demonstrates its promising and intriguing potential as an inspiring platform for exploring the physics of QSL.

\emph{Acknowledgments}---This work was supported by the National Key Research and Development Program of China (Grant No. 2017YFA0302904), the National Science Foundation of China (Grant Nos. U1932215 and No. 11774419), and the Strategic Priority Research Program of the Chinese Academy of Sciences (Grant No. XDB33010100). Q.M.Z. acknowledges the support from Synergetic Extreme Condition User Facility (SECUF), CAS and Users with Excellence Program of Hefei Science Center and High Magnetic Field Facility, CAS. {We thank the ISIS Facility for beam time, RB1910266 (DOI: 10.5286/ISIS.E.RB1910266)\cite{DrJieMa}.} The processing of the ZF-$\mu$SR and LF-$\mu$SR data in Fig. 2 and Fig.3 are based on the MANTID\cite{Arnold2014} software. The calculations and fitting of spin dynamics in Tab. I are based on MathWorks MATLAB software (Academic License for Renmin University of China).

%

\end{document}